\documentclass[aps,pre,twocolumn,showpacs,preprintnumbers,superscriptaddress]{revtex4}

\begin{document}

\draft
\title{Microscopic formula for transport coefficients of causal hydrodynamics}
\author{T.~Koide}
\address{Instituto de F\'{\i}sica, Universidade Federal do Rio de Janeiro, C. P.
68528, 21945-970, Rio de Janeiro, Brazil}
%\ead{koide@fma.if.usp.br}

\begin{abstract}
The Green-Kubo-Nakano formula 
should be modified in relativistic hydrodynamics, 
because of the problem of acausality and the breaking of sum rules.
In this Rapid Communication, 
we propose a formula to calculate the transport coefficients 
of causal hydrodynamics based on the projection operator method.
As concrete examples, we derive the expressions for the diffusion coefficient, 
the shear viscosity coefficient, and corresponding relaxation times.
\end{abstract}

\maketitle

The hydrodynamic model has been applied to 
analyze collective aspects of relativistic heavy-ion collisions.
The analysis, so far, has been implemented extensively for ideal fluid.
However, there is no theoretical reason that we can ignore the effect 
of dissipation.
The relativistic extension of dissipative hydrodynamics is not trivial 
\cite{muronga,dkkm}.
Because of the problem of acausality and the breaking of sum rules, 
the simple covariant extension of the Navier-Stokes equation does not work.
To obtain a consistent theory, we need to introduce a memory effect with 
a relaxation time \cite{dkkm,jou1,gavin}.
Then, the diffusion-type equation is modified to the telegraph-type equation.

This modification also affects the definition of the transport coefficients.
In the usual non-relativistic hydrodynamics, 
it is known that such transport coefficients can be 
calculated by using the Green-Kubo-Nakano (GKN) formula \cite{kubo}.
However, in relativistic hydrodynamics, 
there is a problem of using the GKN formula to estimate the 
transport coefficients of causal hydrodynamics.
For example, let us consider the diffusion process which 
is described by 
\begin{eqnarray}
\frac{\partial}{\partial t}n(x,t) = D_{ac}\nabla^2 n(x,t). \label{ac-de}
\end{eqnarray}
Then, by using the GKN formula, the diffusion constant is given by 
\cite{kubo}
\begin{eqnarray}
D_{ac} 
&=& \left(\frac{\partial \mu}{\partial n} \right)_{\beta}
\lim_{\omega\rightarrow 0}\lim_{{\bf k}\rightarrow 0}\int^{\infty}_0 dt \int d^3 x 
e^{-i({\bf kx}-\omega t)} \nonumber \\
&&\times \frac{1}{3}\int^{\beta}_0 d\lambda 
\langle {\vec j}({\bf x},t-i\lambda){\vec j}({\bf 0},0) \rangle_{eq}, \label{kn}
\end{eqnarray}
where $\mu$ represents a chemical potential and 
$\langle ~~\rangle_{eq}$ denotes the equilibrium expectation value with 
the density matrix $\rho = e^{-\beta H}/Z$ using $Z = Tr[ e^{-\beta H} ]$.
Here, ${\vec j}({\bf x},t)$ is a current operator associated with $n({\bf x},t)$ 
(see Eq. (\ref{ddx}) below). 
In deriving this expression, we assumed that the coarse-grained dynamics 
of $n(x,t)$ is given by Eq. (\ref{ac-de}).
If the coarse-grained dynamics is changed, we cannot use Eq. (\ref{kn}).
Thus the GKN formula should be modified.

As a matter of fact, 
it is known that the diffusion equation does not obey relativistic causality: 
the propagation speed of signal of Eq. (\ref{ac-de}) 
exceeds the speed of light \cite{jou1,gavin,dkkm}. 
Moreover, the diffusion equation is inconsistent 
with sum rules associated with conservation laws 
\cite{kadanoff,koide}.
Aiming to avoid these problems, a frequently considered alternative is
given by the telegraph equation,
\begin{eqnarray}
\frac{\partial^2}{\partial t^2}n({\bf x},t) + \frac{1}{\tau} \frac{\partial}{\partial t}n({\bf x},t)
= D_{c}\nabla^2 n({\bf x},t) .\label{c-de}
\end{eqnarray}
Here, the parameter $\tau$ is the relaxation time and it is necessary to 
obtain a causal theory which is consistent with sum rules \cite{kadanoff,koide}.
When the coarse-grained dynamics is given 
by Eq. (\ref{c-de}) instead of Eq. (\ref{ac-de}), 
the causal diffusion coefficient $D_c$ is not given by Eq. (\ref{kn}).
It should be noted that 
the telegraph equation will not be a unique solution to solve the problem of acausality.
See \cite{footnote1} for details.

In Rapid Communication, we derive a general formula of 
transport coefficients of causal hydrodynamics.
For example, as we will see later, in the case of the causal diffusion process described by Eq. (\ref{c-de}), 
the causal diffusion coefficient and the relaxation time are 
given by Eqs. (\ref{cdc}) and (\ref{rt}).
Interestingly, 
the diffusion constant (\ref{kn}) does not represent the causal diffusion constant.
It reduces to the inverse of the relaxation time under some approximation.

First of all, we consider an operator of a 
gross (hydrodynamic) variable $O({\bf x})$.
The time-evolution of the $O$ is given by the Heisenberg equation of motion,
\begin{eqnarray}
\frac{\partial}{\partial t}O({\bf x},t) = iLO({\bf x},t),
\end{eqnarray}
where $L$ is the Liouville operator defined by $LA = [H,A]$.
Here, $A$ is an arbitrary operator and $H$ is the Hamiltonian of our system.
The operator $O$ is arbitrary, but should satisfy 
$\langle O \rangle_{eq} = 0$ \cite{km,koide}.
The Heisenberg equation of motion contains the information not only of 
gross variables associated with hydrodynamic time scales, but also of 
microscopic variables with shorter time scales.
To carry out coarse grainings of the latter irrelevant information, 
we introduce a projection operator $P$ and its complementary $Q=1-P$ 
\cite{nakajima,mori,shibata,km2,km,koide}.
Then, the Heisenberg equation of motion is rewritten as
\begin{eqnarray}
\frac{\partial}{\partial t}O({\bf x},t) 
&=& e^{iLt}PiLO({\bf x})  \nonumber \\
&&\hspace{-2.5cm} + \int^{t}_{0}ds e^{iL(t-s)}PiLQe^{iLQs}iLO({\bf x}) 
+ Qe^{iLQt}iLO({\bf x}). \label{tc}
\end{eqnarray}
This equation is called the time-convolution equation 
\cite{shibata,km2}.
The first term on the right-hand side (rhs) is called the streaming term and corresponds to 
a collective oscillation which does not produce entropy.
The second term represents the dissipative part of the time evolution.
The third term is interpreted as the noise term.
The irreversible current is described by the second term.
Irreversible currents are proportional to thermodynamic forces and 
the transport coefficients are defined by the proportional coefficients.
Thus the first and third terms, which 
are irrelevant to the definition of transport coefficients, 
are ignored in the following discussion.

So far, we did not specify the form of the projection operator.
In this work, we use the Mori projection operator \cite{mori}, 
\begin{eqnarray}
PA = \int d^3 x d^3 x' (A, O^{\dagger}({\bf x})) (O({\bf x}),O^{\dagger}({\bf x}'))^{-1} O({\bf x'}),
\end{eqnarray}
where the Kubo's canonical correlation is defined by
\begin{eqnarray}
(A,B) 
&=& 
\int^{\beta}_{0}\frac{d\lambda}{\beta}
\langle e^{\lambda H}Ae^{-\lambda H}B \rangle_{eq}.
\end{eqnarray}
For the physical meaning of the Mori projection, see \cite{mori,km}.
Substituting the Mori projection into Eq. (\ref{tc}), 
we have
\begin{eqnarray}
\frac{\partial}{\partial t}O ({\bf x},t) 
= -\int^{t}_{0}ds \int d^3 x' \Xi({\bf x},{\bf x}';s)O({\bf x}',t-s). \label{tc2}
\end{eqnarray}
The memory function $\Xi$ is given by 
\begin{eqnarray}
\Xi({\bf x},{\bf x}'';t)
&=& -\theta(t) \int d^3 x' \nonumber \\
&&\hspace{-2cm} \times (iLQe^{iLQt}iLO({\bf x}),O^{\dagger}({\bf x}')) 
(O({\bf x}'),O^{\dagger}({\bf x}''))^{-1}. \label{memory}
\end{eqnarray}
It is difficult to calculate the memory function 
because of the coarse-grained time-evolution operator $e^{iLQt}$.
Usually, we approximately replace this term with the usual time-evolution operator 
$e^{iLt}$. 
As is discussed in \cite{fick}, the approximation was considered to be justified 
at least for the low momentum limit, 
and the diffusion-type equation is derived. 
However, as is shown in \cite{koide}, 
this approximation is not justified even for the low momentum limit, 
and gives rise to several problems: inconsistency of causality and sum rules.
We will come back to this point later.

Recently, the coarse-grained time-evolution operator 
was calculated exactly \cite{sawada,km,koide}.
By using the same method, we find that the Fourier transform of 
the memory function is given by 
\begin{eqnarray}
\Xi^F({\bf k},\omega) 
= -\frac{\ddot{X}^F({\bf k},\omega)}{1 + \dot{X}^F({\bf k},\omega)}
+ \frac{\dot{X}^F({\bf k},t=0)\dot{X}^F({\bf k},\omega)}{1+\dot{X}^F({\bf k},\omega)}. \label{xi}
\end{eqnarray}
Here, $\dot{X}^F({\bf k},\omega)$ and $\ddot{X}^F({\bf k},\omega)$ 
are given by the Fourier transforms of the 
following correlation functions, respectively,
\begin{eqnarray}
\dot{X}({\bf x}-{\bf x}'',t)
&=& -\frac{i}{\beta}\theta(t)\int d^3 x' 
\langle [O({\bf x},t),O^{\dagger}({\bf x}'])
\rangle_{eq} 
\nonumber \\
&&\times (O({\bf x}'),O^{\dagger}({\bf x}''))^{-1},\\
\ddot{X}({\bf x}-{\bf x}'',t)
&=& -\frac{i}{\beta}\theta(t)\int d^3 x' 
\langle [iL O({\bf x},t),O^{\dagger}({\bf x}')]\rangle_{eq} \nonumber \\
&&\times (O({\bf x}'),O^{\dagger}({\bf x}''))^{-1}.
\end{eqnarray}
This is the first main result of this work.
A similar expression of the memory function is obtained 
for the current-current correlation 
function in \cite{sawada}, although the term that corresponds to 
the second term on the rhs of Eq. (\ref{xi}) does not exist 
because they considered the case of $\dot{X}^F({\bf k},t=0)=0$.

To obtain the expression of transport coefficients, 
we have to employ an approximation to derive a time-convolutionless equation 
from the integrodifferential equation.
Following \cite{km,koide}, we separate the memory function into two terms,
\begin{eqnarray}
\frac{\partial}{\partial t}O({\bf k},t) 
&=& -\int^{t}_{0}d\tau \Omega^2({\bf k},t-\tau) O({\bf k},\tau) \nonumber \\
&&- \int^{t}_{0}d\tau \Phi({\bf k},t-\tau) O({\bf k},\tau),
\end{eqnarray}
where the frequency function and the renormalized memory function are, 
respectively, defined by 
\begin{eqnarray}
\Omega^2({\bf k},t)
&=& i\int \frac{d\omega}{2\pi}{\rm Im}[\Xi^F({\bf k},\omega)]e^{-i\omega t}, \\
\Phi({\bf k},t)
&=& \int \frac{d\omega}{2\pi}{\rm Re}[\Xi^F({\bf k},\omega)]e^{-i\omega t}.
\end{eqnarray}

To employ this approximation, we have to know the temporal behavior of 
the two functions.
So far, the behaviors of the two functions are investigated 
for the chiral order parameter \cite{km} 
and the nonrelativistic number density \cite{koide}.
For these cases, the two functions showed common properties; 
the frequency function converges to a finite value and the 
renormalized memory function vanishes at late time. 
Thus we employ an important assumption for the temporal behavior 
of the two functions. 
The renormalized memory function relaxes rapidly and vanishes at late time.
On the other hand, the frequency function converges to a finite value 
after short time evolution.
Then, the approximation means to replace 
the frequency function and the renormalized memory function 
with the following functions:
\begin{eqnarray}
\Omega^2({\bf k},t)
\longrightarrow
D_{{\bf k}}{\bf k}^2, ~~~~
\Phi ({\bf k},t)
\longrightarrow
\frac{2}{\tau_{\bf k}}\delta(t),
\end{eqnarray}
where
\vspace{-0.5cm}
\begin{eqnarray}
D_{\bf k} 
&=& \frac{1}{{\bf k}^2}\lim_{t\rightarrow \infty}\Omega^2({\bf k},t), \label{cdc} \\
\frac{1}{\tau_{\bf k}}
&=& \int^{\infty}_{0}dt \Phi({\bf k},t). \label{rt}
\end{eqnarray}
These two definitions are the second main result of this Rapid Communication.
The coefficient $D_{\bf k}$ gives us the dissipative coefficient 
and the coefficient $\tau_{\bf k}$ is the relaxation time.
As a matter of fact, by using these expressions, we arrive at 
\begin{eqnarray}
\frac{\partial}{\partial t}O({\bf k},t) + \frac{1}{\tau_{\bf k}}O({\bf k},t)= 
- D_{\bf k}{\bf k}^2\int^{t}_0 d\tau O({\bf k},\tau).  \label{markoveq}
\end{eqnarray}
This equation has the same form as the causal diffusion equation 
(\ref{c-de}) except for the momentum dependence of the relaxation time 
$\tau_{\bf k}$ \cite{koide}.

The same equation as Eq. (\ref{markoveq}) has already been given by \cite{km,koide}.
However, the perturbative approximation was applied in the calculation of the 
memory functions there.
The result of this Rapid Communication is exact except for the local approximation.

The validity of the above assumption for the frequency 
and renormalized memory functions 
should be checked by calculating these temporal behaviors, numerically.
However, 
we can see that the assumption is reasonable by using 
the final value theorem of the Laplace transformation \cite{koide, km}.
When the renormalized memory function converges to zero at late time, 
its Laplace transform should satisfy, 
\begin{eqnarray}
\lim_{t\rightarrow \infty} \Phi({\bf x},t)
=
\lim_{s\rightarrow 0}s\Phi^L({\bf x},s) 
= 0.
\end{eqnarray}
Similarly, the dissipative coefficient which is calculated 
from the frequency function should be given by 
\begin{eqnarray}
D_{\bf k}k^2 
= \lim_{t\rightarrow \infty} \Omega^2({\bf k},t)
= \lim_{s\rightarrow 0}s(\Omega^{2})^L({\bf k},s).
\end{eqnarray}
Here, the suffix $L$ means the Laplace transform.

To obtain the dissipative coefficient and the relaxation time, 
we have to know the correlation functions 
$\dot{X}^F({\bf k},\omega)$ and $\ddot{X}^F ({\bf k},\omega)$.
Both of them can be calculated by using the imaginary-time Green function,
${\cal G}({\bf x},\tau;{\bf x}',\tau') 
= - {\rm Tr}\{ \rho T_{\tau}O({\bf x},\tau)O^{\dagger}({\bf x},\tau') \}$.
Once we have it, the correlation functions $\dot{X}^F({\bf k},\omega)$, 
which is essentially the retarded Green function, 
can be estimated by using the analytic continuation of 
the imaginary-time Green function.
The other correlation function $\ddot{X}^F({\bf k},\omega)$ can be calculated 
by using the following relation,:
\begin{eqnarray}
\frac{\partial}{\partial t}\dot{X}({\bf x}-{\bf x}'';t)
= \ddot{X}({\bf x}-{\bf x}'';t)
+ \dot{X}({\bf x}-{\bf x}'';t)\delta(t).
\end{eqnarray}

To see the relation between our formula and the GKN formula, 
we would like to discuss the diffusion process of a conserved number density 
$n({\bf x},t)$, 
by setting $O = \delta n({\bf x}) = n({\bf x}) - \langle n({\bf x}) \rangle_{eq}$. 
When we approximately replace $e^{iLQt}$ of Eq. (\ref{memory}) with $e^{iLt}$ 
as was mentioned below Eq. (\ref{memory}),  
the memory function is given by 
\begin{eqnarray}
\Xi^F({\bf k},\omega)
\approx - \ddot{X}^F({\bf k},\omega).\label{nocg}
%+ \dot{X}({\bf k},0)\dot{X}^F({\bf k},\omega). 
\end{eqnarray}
Here, we used $\dot{X}^F({\bf k},t=0) = 0$ for $O = \delta n({\bf x})$.
The correlation function $ \ddot{X}^F({\bf k},t)$ 
is 
\begin{eqnarray}
\ddot{X}^F({\bf k},t)
&=& \theta(t) \beta \left( \frac{\partial \mu}{\partial n} \right)_{\beta} 
\nonumber \\
&&\times \int d^3 x 
e^{-i{\bf kx}}\frac{1}{3}k^2 ({\vec j}({\bf x},t),{\vec j}({\bf 0},0)), 
\label{ddx}
\end{eqnarray}
where $\mu$ is the chemical potential.
Here, we introduced the current operator by 
$d n({\bf x},t)/dt = - \nabla {\vec j}({\bf x},t)$.
And the equal-time correlation function is approximately given by 
$\int d^3 x e^{-i{\bf kx}} (\delta n({\bf x},0),\delta n({\bf 0},0)) 
\approx \frac{1}{\beta}\left( \frac{\partial n}{\partial \mu} \right)_{\beta}$ 
\cite{kubo}.
At the low momentum limit, we have
\begin{eqnarray}
D_{\bf k} 
= 0, ~~~~
\frac{1}{\tau_{\bf k}}
= D_{ac}k^2,
\end{eqnarray}
where $D_{ac}$ is defined by Eq. (\ref{kn}).
Then, one can see that Eq. (\ref{markoveq}) is reduced to the diffusion equation (\ref{ac-de}), and 
the diffusion constant is given by Eq. (\ref{kn}).
From this analysis, we found that our formula is reduced to the GKN formula 
when we ignore the effect of coarse grainings on correlation functions.
Then, the GKN formula gives not the diffusion coefficient $D_{c}$ but the 
inverse of the relaxation time $1/\tau_{\bf k}$.

It should be noted that 
the diffusion equation cannot satisfy 
the f-sum rule.
In general, when we take the effect of the coarse-grained 
time-evolution operator into account, $D_{\bf k}$ does not vanish and we obtain 
the telegraph equation.
The telegraph equation can be consistent with the f-sum rule 
\cite{kadanoff,koide}.

Next, we apply our formula to a nonrelativistic fluid to derive 
the shear viscosity constant.
In the classical calculation, the commutator should be replaced by 
the Poisson bracket.
For simplicity, we consider a particular case of shear flow, where 
the fluid velocity $u_x(y,t)$ points in the $x$ direction and varies 
spatially only in the $y$ direction.
Then, the current density is given by 
$J^x(k_y) = \rho u_x= \sum p^x_i e^{ik_y y_i}$,
where $\rho$ is a mass density and $p^x_i$ is a momentum of a constituent particle.
By setting $O=J^x(k_y)$ and substituting into Eq. (\ref{markoveq}), 
we have the evolution equation of the current density, 
\begin{eqnarray}
\frac{\partial}{\partial t}J^x(k_y,t) + \frac{1}{\tau_{k_y}}J^x(k_y,t)
&=& - D_{k_y}k^2_{y}\int^t_0 d\tau \rho u^x(k_y,\tau).\nonumber \\
\end{eqnarray}
If we ignore the effect of the coarse graining,  
we obtain, 
\begin{eqnarray}
D_{k} = 0,~\frac{1}{\tau_{k}} = k^2 \frac{\beta}{\rho V}
\int^{\infty}_{0}dt \langle P_{xy}(k,t)P_{xy}(-k,0) \rangle_{eq},
\end{eqnarray} 
where the pressure tensor is defined by $\dot{J}^x = ik_y P_{xy}(k_y)$ and 
$V$ is the total volume of the system.
At the low momentum limit, the inverse of the relaxation time is reduced to 
the GKN formula for the shear viscosity constant \cite{zwanzig-b}.

Similarly, we can derive the formula for the shear viscosity of 
the relativistic dissipative hydrodynamics \cite{dkkm}.
We consider a particular situation which is the same as the nonrelativistic case 
discussed above.
Then, the energy-momentum tensor $T^{\mu\nu}$ 
obeys the following equation of motion:
$\frac{\partial}{\partial t} T^{0x} = - \frac{\partial}{\partial y} T^{yx} 
= - \frac{\partial}{\partial y} \pi^{xy}$,
where $\pi^{\mu\nu}$ is the shear viscosity.
Here, we ignore the reversible part.
As is discussed in \cite{dkkm}, the shear viscosity is given by solving the 
telegraph-type equation.
Then, the linearized equation of motion in terms of the fluid velocity is 
\begin{eqnarray}
\frac{\partial^2}{\partial t^2}T^{0x}
+ \frac{1}{\tau} \frac{\partial}{\partial t}T^{0x}
+ \frac{\eta}{2\tau} \frac{\partial^2}{\partial y^2} u^x = 0. \label{eqvis}
\end{eqnarray}
On the other hand, by setting $O = T^{0x}$, we obtain 
\begin{eqnarray}
&&\frac{\partial}{\partial t}T^{0x}(k_y,t) + \frac{1}{\tau_{k_y}}T^{0x}(k_y,t) \nonumber \\
&&\hspace{2cm}= -D_{k_y}k^2_y \int^{t}_{0}d\tau w u^x(k_y,\tau),
\end{eqnarray}
where $w$ is an enthalpy density.
Here, we assume that the fluid velocity is given by 
$ u^x(k_y) = T^{0x}(k_y)/w$ in the linear approximation of the fluid velocity.
Comparing with Eq.~(\ref{eqvis}), 
we obtain the expressions of the transport coefficients of relativistic 
dissipative hydrodynamics. 
The relaxation time and the shear viscosity constant 
are given by 
\begin{eqnarray}
\tau &=& \lim_{k_y \rightarrow 0} \tau_{k_y}, \\
\eta &=& \lim_{k_y \rightarrow 0} 2 w D_{k_y} \tau_{k_y},
\end{eqnarray}
respectively.
In particular, the correlation function $\ddot{X}({\bf k},t)$ is given by
\begin{eqnarray}
\ddot{X}^F ({\bf k},t)
&=& \theta(t)  
\frac{1}{5}\int d^3 x d^3 x_1
e^{-i{\bf kx}} k^2 (T^{\alpha\beta}({\bf x},t),T_{\alpha\beta}({\bf x}_1,0))
\nonumber \\
&&\hspace{-1cm}\times (T^{0x}({\bf x}_1,0),T^{0x}({\bf 0},0))^{-1}~~
(\alpha,\beta=1,2,3).
\end{eqnarray}
If we ignore the effect of the coarse-grainings again, our calculation reproduces the 
shear viscosity given by \cite{hosoya}.
In this sense, our expression of the shear viscosity is the generalization of 
the well-known result.

In this work, we derived a formula to calculate transport coefficients 
of causal hydrodynamics.
In our formula, we took the effect of the coarse grainings in the memory function 
into account.
Then, the coarse-grained equation is given not by the diffusion-type equation, 
but by the telegraph-type equation.
By comparing our equation with phenomenological equations, 
we obtained the expressions of the 
diffusion coefficient, the shear viscosity coefficient and the relaxation times.
This formula reproduces the GKN formula when 
the effect of the coarse grainings is ignored in the memory function.

Near phase transitions, the relaxation of 
the fluctuations of thermodynamic quantities 
is decelerated. 
This is called the critical slowing down 
and is expected to happen for the shear viscosity \cite{mclerran}. 
We have already showed that the critical slowing down near 
the chiral phase transition can be described in this formalism \cite{km}.

In this Rapid Communication, we derived the linear equation (\ref{tc2}) 
and ignored the effect of nonlinearity 
included in the noise term, the third term of 
Eq. (\ref{tc}), which is negligibly small near equilibrium.
When we discuss the far-from-equilibrium dynamics, 
which will be described by the nonlinear hydrodynamics, we have to extend the 
definition of the projection operator so as to extract the non-linear effect from 
the noise term \cite{mori,km}.

The formula of transport coefficients of 
extended thermodynamics is discussed also by Ichiyanagi \cite{ichi}.
The formula, however, contains the coarse-grained time-evolution operator, 
explicitly.
Thus to apply the formula to practical problems, 
we need to rewrite the expression as was done in this work.

The author thanks G. S. Denicol, F. Gelis, T. Kodama, and Ph. Mota for helpful discussions.
This work was supported by CNPq.


\begin{thebibliography}{99}
%
\bibitem{muronga}
A. Muronga, Phys. Rev. Lett. {\bf 88}, 062302 (2002); 
{\bf 89}, 159901(E) (2002);
Phys. Rev. \textbf{C69}, 034903 (2004);
nucl-th/0611090;
U. Heinz, H. Song, and A.K. Chaudhuri, Phys. Rev. \textbf{C73}, 034904 (2006);
A. K. Chaudhuri, arXiv:0704.0134;
R. Baier, P. Romatschke, and U. A. Wiedemann, 
Phys. Rev. \textbf{C73}, 064903 (2006); 
R. Baier and P. Romatschke, nucl-th/0610108;
P. Romatschke, nucl-th/0701032.
%
\bibitem{dkkm}
T. Koide, G. S. Denicol, Ph. Mota, and T. Kodama, Phys. Rev. {\bf C75}, 034909 (2007) .
%
\bibitem{jou1}
D.~Jou, J.~Casas-V\'azquez, and G.~Lebon, 
Rep.~Prog.~Phys.~{\bf 51}, 1105 (1988); {\bf 62}, 1035 (1999);
I. M\"uller, Living Rev. Relativ. {\bf 2}, 1 (1999).
%
\bibitem{gavin}
M.~Abdel-Aziz and S.~Gavin, Phys. Rev. {\bf C70}, 034905 (2004);
T. Koide, G. Krein, and R. O. Ramos, Phys. Lett. {\bf B636}, 96 (2006).
%
\bibitem{kubo}
R. Kubo, M. Toda, and N. Hashitsume, 
{\it Statistical Physics II} (Springer-Verlag, Berlin, 1983).
%
\bibitem{kadanoff}
L. P. Kadanoff and P. C. Martin, Ann. Phys. (N.Y.) {\bf 24}, 419 (1963).
%
\bibitem{koide}
T. Koide, Phys. Rev. E{\bf 72}, 026135 (2005).
%
\bibitem{footnote1}
W. L. Kath, Physica {\bf D12}, 375 (1984); 
J. Dunkel, P. Talkner, and P. H\"anggi, Phys. Rev. {\bf D75}, 043001 (2007).
%
\bibitem{km}
T. Koide and M. Maruyama, Nucl. Phys. {\bf A742}, 95 (2004).
%
\bibitem{nakajima}
S. Nakajima, Prog. Theor. Phys. {\bf 20}, 948 (1958);
R. Zwanzig, J. Chem. Phys. {\bf 33}, 1338 (1960).
%
\bibitem{mori}
H. Mori, Prog. Theor. Phys. {\bf 33}, 423 (1965).
%
\bibitem{shibata}
N. Hashitaume, F. Shibata, and M. Shing\~u, J. Stat. Phys. {\bf 17}, 155 (1977);
F. Shibata, Y. Takahashi, and N. Hashitaume, {\it ibid}. {\bf 17}, 171 (1977);
F.~Shibata and T.~Arimitsu, J.~Phys.~Soc.~Jpn~{\bf 49}, 891 (1980);
C.~Uchiyama and F.~Shibata, Phys.~Rev.~{\bf E60}, 2636 (1999).
kada%
\bibitem{km2}
T. Koide and M. Maruyama, Prog. Theor. Phys. {\bf 104}, 575 (2000);
T. Koide, {\it ibid}. {\bf 107}, 525 (2002).
%
\bibitem{sawada}
J. Okada, I. Sawada, and Y. Kuroda, J. Phys. Soc. Jpn. {\bf 64}, 4092 (1995).
%
\bibitem{fick}
E. Fick and G. Sauermann, {\it The Quantum Statistics of Dynamic Process} 
(Springer-Verlag, Berlin, 1983).
%
\bibitem{zwanzig-b}
R. Zwanzig, {\it Nonequilibrium Statistical Mechanics} 
(Oxford University, New York, 2004).
%
\bibitem{hosoya}
A. Hosoya, M. Sakagami, and M. Takao, Ann. Phys. (N.Y.) {\bf 154}, 229 (1984).
%
\bibitem{mclerran}
L. P. Csernai, J. I. Kapusta, and L. D. Mclerran, 
Phys. Rev. Lett. {\bf 97}, 152303 (2006).
%
\bibitem{ichi}
M. Ichiyanagi, Prog. Theor. Phys. {\bf 84}, 810 (1990).
%
\end{thebibliography}
\end{document}